\documentclass[conference]{IEEEtran}
\usepackage{textcomp}

   \usepackage[pdftex]{graphicx}
   \DeclareGraphicsExtensions{.pdf, .jpeg}
\begin{document}
\title{Analysis of Short-Term Stability of Miniature \textsuperscript{171}Yb\textsuperscript{+} Buffer Gas Cooled Trapped Ion Clock}

\author{\IEEEauthorblockN{David R. Scherer, C. Daniel Boschen, Jay Noble, Michael Silveira, Dwayne Taylor,\\ Jonathan Tallant, K. Richard Overstreet, S. R. Stein}
\IEEEauthorblockA{Microsemi\\
Beverly, MA 01915\\}}

%

\maketitle

\begin{abstract}
We demonstrate an improvement in short-term \mbox{stability} by a factor of 10 over a previous generation miniature buffer gas cooled trapped ion clock. We describe the enhancement to detection \textit{SNR} that has enabled this improvement, the method of clock operation, and the measurement of clock short-term stability. Additionally, we numerically investigate the magnitude of the Dick effect in our pulsed ion clock.
\end{abstract}

%
\IEEEpeerreviewmaketitle

\section{Introduction}
Several approaches have been pursued in the development of miniature atomic clocks in systems with a physics package smaller than 100~cm\textsuperscript{3}. Within the context of the DARPA IMPACT program, these have included cold-atom clocks \cite{honeywell, scherer}, optical clocks \cite{oewaves}, and trapped-ion clocks \cite{IMPACTAPL, IMPACT_fcs}. All of these systems have demonstrated a short-term stability of $\sigma_y(\tau) = 1 \cdot 10^{-11} \tau^{-1/2}$. Here we improve upon the short-term stability of the trapped-ion approach by an order of magnitude, demonstrating a short-term stability of $\sigma_y(\tau) = 1.6 \cdot 10^{-12} \tau^{-1/2}$ in the same miniature vacuum package, but using laboratory-scale lasers, optics, and electronics. 

\section{Vacuum Package and Experimental Control}
\begin{figure}
\begin{center}
\leavevmode
\includegraphics[width=0.7\linewidth]{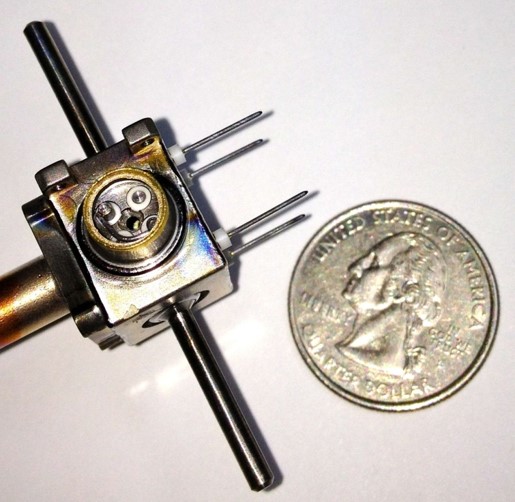}
\end{center}
\caption{Photograph of miniature trapped-ion vacuum package developed by JPL. Two external Yb ovens are shown on the top and bottom, the copper pump-out tube is on the left, and four electrical feedthroughs for connections to the trap rods are on the right. Four trap rods surrounded by insulating material are visible through the sapphire window. In the middle of the window, a small aperture allows for optical access for the laser beams. The collection window (not visible) is located opposite the pump-out port, in between the four electrical feedthroughs.} \label{fig:photo-trap}
\end{figure}

\begin{figure}
\begin{center}
\leavevmode
\includegraphics[width=1\linewidth]{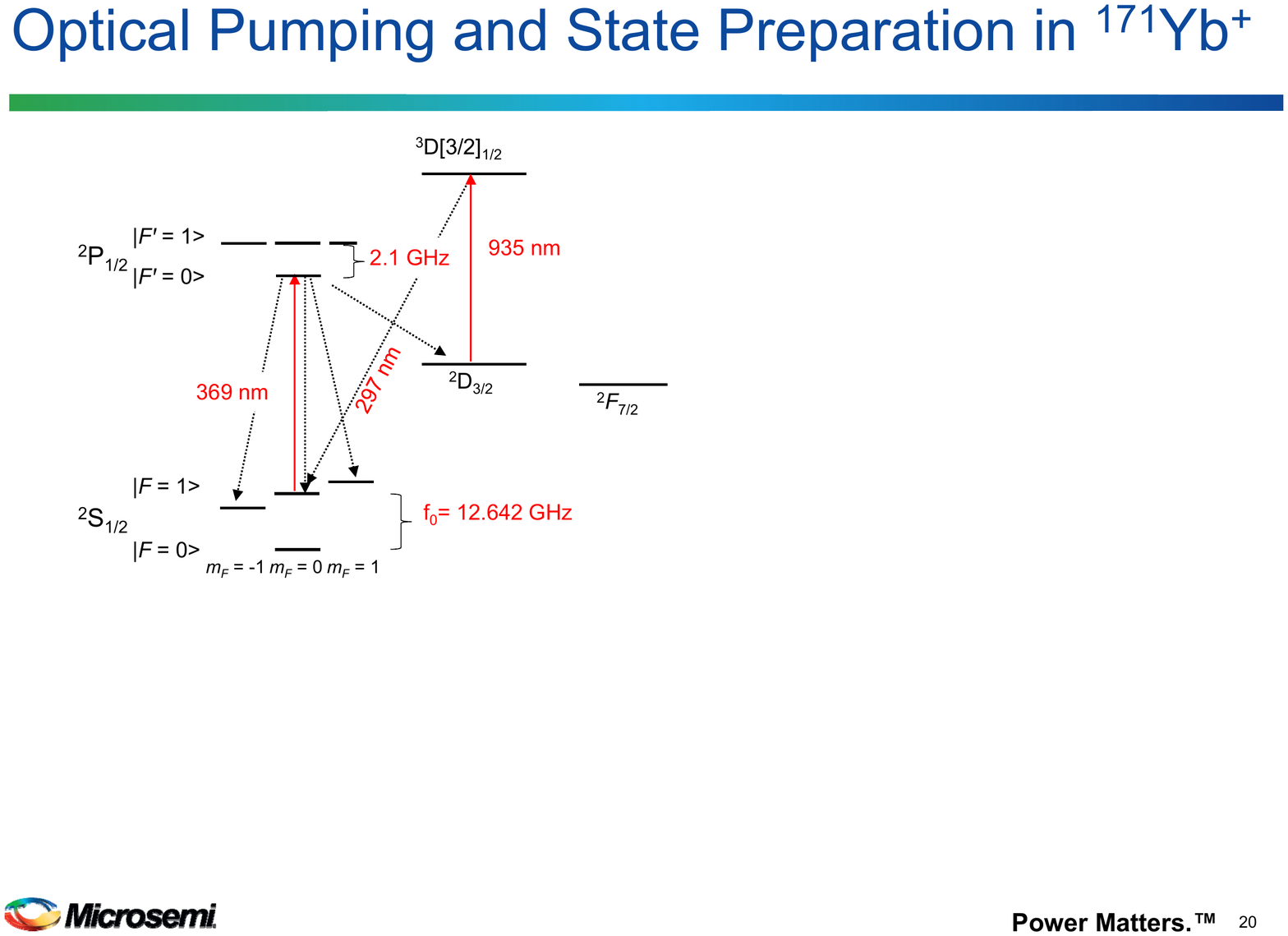}
\end{center}
\caption{Energy levels of \textsuperscript{171}Yb\textsuperscript{+}. The clock transition is between the $\mid$$F$=0, $m\textsubscript{F}$=0$>$ $\rightarrow$ $\mid$$F$=1, $m\textsubscript{F}$=0$>$ transition at 12.642 GHz. Optical pumping and state readout is accomplished via the 369 nm laser. The D-state is continuously cleared with a 935 nm laser.}
\label{fig:171-energy-levels}
\end{figure}

We use a vacuum package nearly identical to that developed by Jet Propulsion Laboratory and used previously by Sandia National Laboratory \cite{IMPACTAPL}, shown in Figure~\ref{fig:photo-trap}. The vacuum package consists of a titanium chamber with brazed sapphire windows and two Yb appendage ovens, one filled with natural Yb and one with isotopically-enriched \textsuperscript{171}Yb. As in previous iterations, the vacuum package contains a non-evaporable getter pump near the copper pump-out tube. The only significant difference between this vacuum package and the previous effort is a slightly larger (7.9~mm inner diameter) copper pump-out port, which allows for larger conductance to the main vacuum pump. The previous effort \cite{IMPACT_fcs} was focused on miniaturization and used a compact micro-optical light collection system at the 297~nm detection pathway due to laser scattering at the probe wavelength of 369~nm (Fig. \ref{fig:171-energy-levels}). In contrast, our effort, focused on extending the performance of the miniature vacuum package, utilizes macroscopic lenses and collects ion fluorescence at 369~nm.

The vacuum package is connected via a 15-cm long copper tube to a chamber that is actively pumped by a 250~L/s turbo pump and contains an inlet for Ne buffer gas. In the vicinity of the turbo pump the base pressure of the chamber is \mbox{1 $\cdot$ 10\textsuperscript{-10}~Torr}.  

The lasers are frequency stabilized to a High Finesse wavemeter at the wavelengths indicated in Figure~\ref{fig:171-energy-levels}. A Toptica External Cavity Diode Laser (ECDL) laser system at 739~nm is frequency-doubled in a resonant cavity to generate the 369~nm light used for optical pumping and state readout. A weak probe beam of 50~$\mu$W is used for state readout, while a higher-power (2~mW) beam is used for trap loading. The 369~nm laser is stabilized to a wavelength corresponding to the \textsuperscript{2}S\textsubscript{1/2} $\mid$$F$=1, $m\textsubscript{F}$=0$>$ $\rightarrow$ \textsuperscript{2}P\textsubscript{1/2}$\mid$$F$=0, $m\textsubscript{F}$=0$>$ transition. A Toptica ECDL laser system at 399~nm is used for isotope selective two-photon photoionization, with a power of 1~mW. A DFB laser at 935~nm of power 4~mW is used for clearing the D-state. The D-state clearing laser is stabilized at the \textsuperscript{2}D\textsubscript{3/2} $\mid$$F$=1, $m\textsubscript{F}$=0$>$ $\rightarrow$ \textsuperscript{3}D[3/2]\textsubscript{1/2}$\mid$$F$=0, $m\textsubscript{F}$=0$>$ transition. The UV beams are combined and co-propagated with the IR beam and pass through the long axis of the ion trap. The UV beams are weakly focused through the vacuum package, with a beam diameter of approximately 100~$\mu$m.


Microwave interrogation is achieved with a custom microwave synthesizer at 12.6~GHz. The synthesizer is based on a Morion MV197 oven-controlled quartz oscillator (OCXO) with a short-term stability of \mbox{1.5 $\cdot$ 10\textsuperscript{-12}} at 1 sec. For closed-loop clock operation, the synthesizer is controlled with a tuning voltage based on an error signal computed after microwave interrogation. The magnetic bias C-field is approximately 440~mG. The vacuum package is surrounded by a single-layer mu-metal shield. The laser shutters, microwave field, and photon counter gate are controlled by a computer.

The trap is a linear quadrupole trap \cite{prestage_linear} operated in two-rod configuration, with two opposing trap rods driven at identical RF potential. Instead of applying a positive DC bias voltage to the endcaps, the endcaps and chamber body are grounded, while all four rods experience a -20~V DC bias potential. The 12.6~GHz microwave field is coupled to the two trap rods that do not carry RF. The rod spacing (center-to-center) is 3.23~mm, the rod length is 8.89~mm, and the rod diameter is 1.52~mm. The ratio of the rod radius to the distance between the rod center and the nodal line is 0.33. While a ratio of 0.53 is optimum in terms of minimal perturbations to an ideal quadrupole field \cite{janik}, this would block too much optical access for light collection.

\begin{figure}
\begin{center}
\leavevmode
\includegraphics[width=1\linewidth]{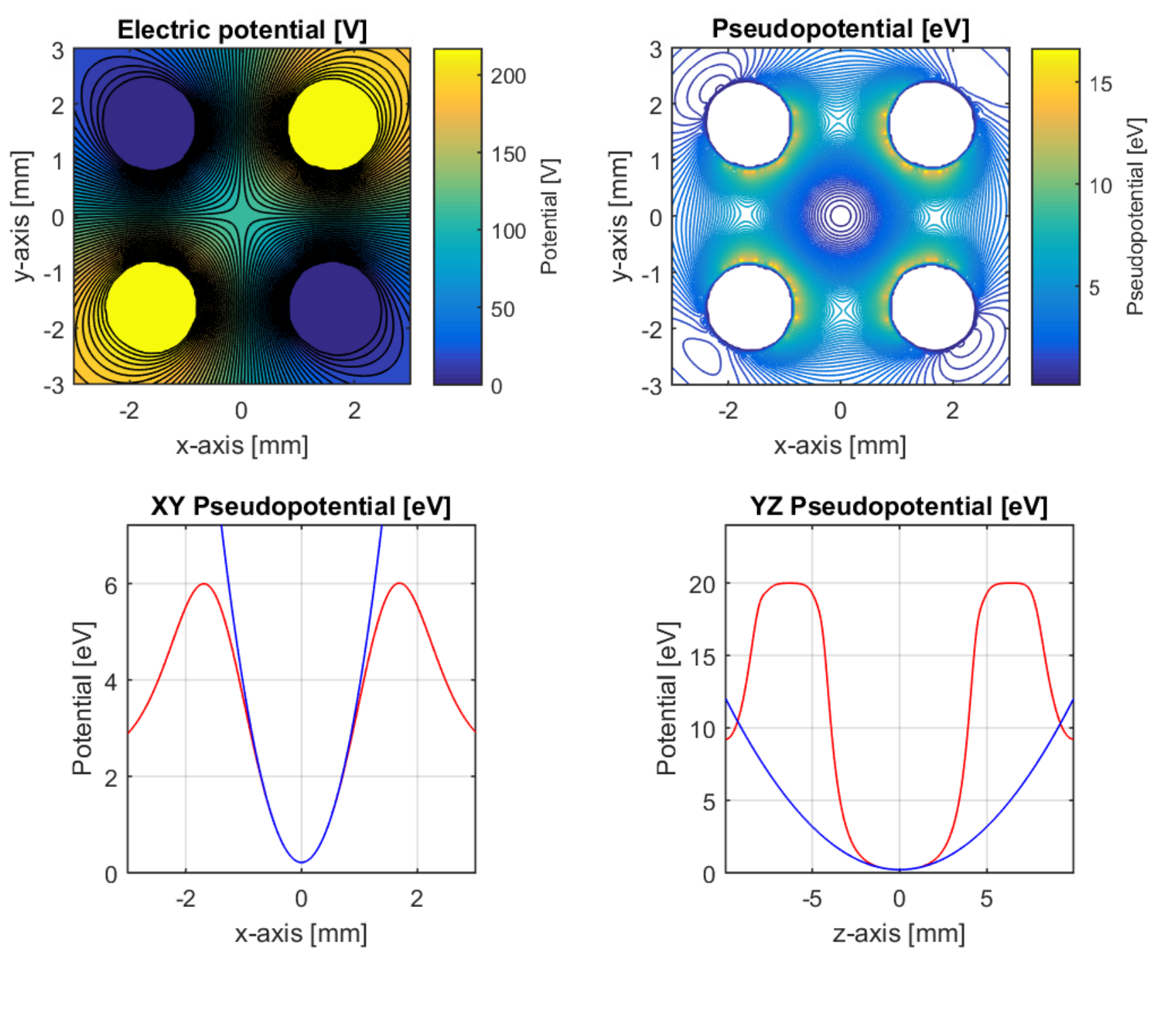}
\end{center}
\caption{Electric potential in the radial plane at the center of the trapping region. Pseudopotential in the radial plane at the center of the trapping region. Pseudopotential (red) along a line in the radial (XY) plane at the center of the trapping region in between the trap rods, along with harmonic fit (blue) corresponding to the computed secular frequency. Pseudopotential (red) along the nodal line in the axial (YZ) plane, showing poor fit (blue) to a harmonic potential.} \label{fig:pseudopotential}
\end{figure}

We compute the instantaneous electric potential by solving the Laplace equation using a charged particle solver to determine the electric potential $V(t)$ throughout the trapping region. We then compute the time-averaged effective pseudopotential $\Psi(t)$ using the pseudopotential approximation.

\begin{equation}
\Psi(x,y,z) = \frac{q E^2(x,y,z)}{4 m \Omega^2} = \frac{q}{4 m \Omega^2}|\nabla V(x,y,y)|^2
\end{equation}

\begin{figure}
\begin{center}
\leavevmode
\includegraphics[width=1\linewidth]{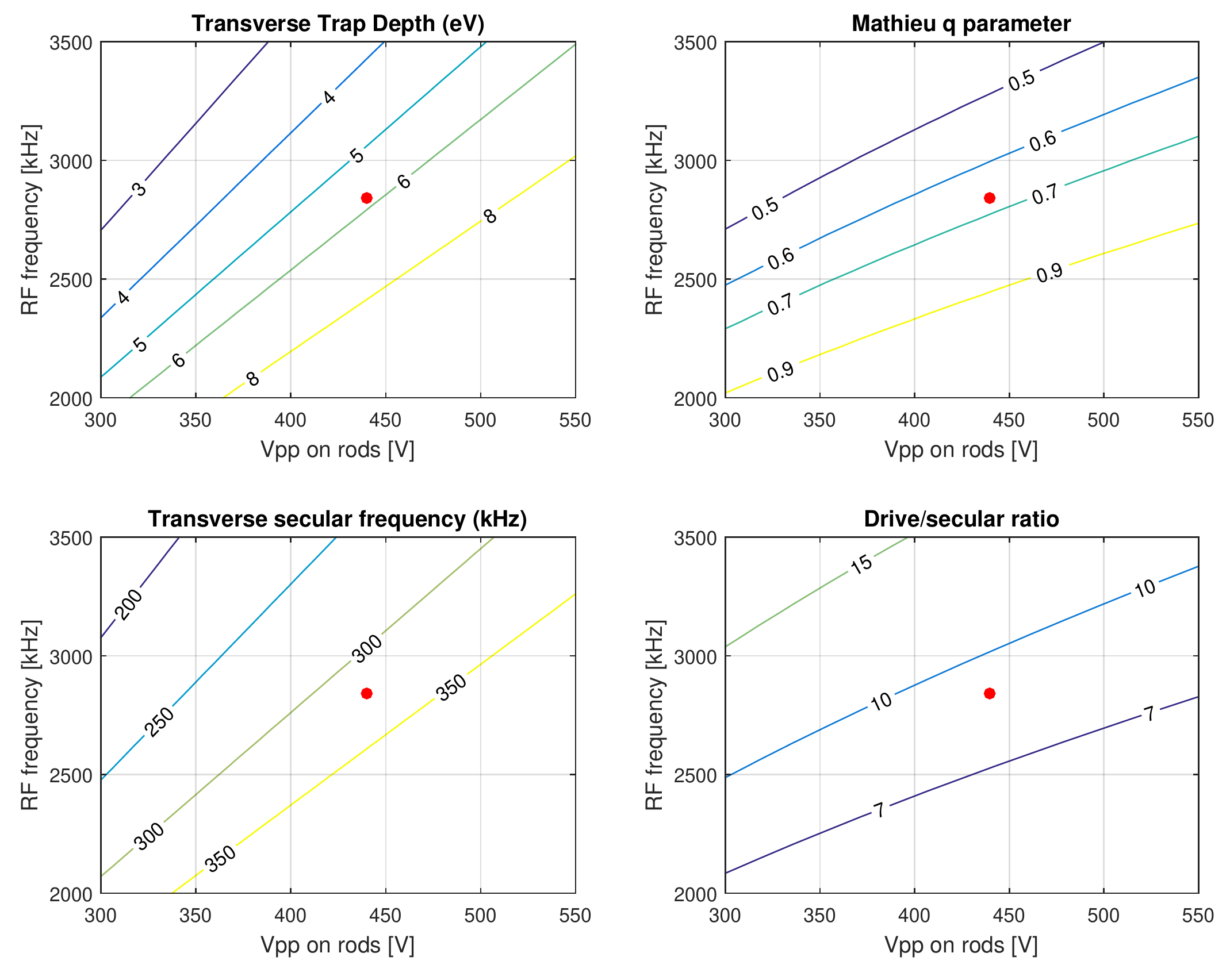}
\end{center}
\caption{Trap depth (eV), Mathieu $q$ parameter, transverse secular frequency (kHz), and drive frequency to secular frequency ratio as a function of V\textsubscript{pp} on the rods and RF drive frequency. The operating point is highlighted with a red dot.} \label{fig:contour}
\end{figure}

We operate the trap at 440~V\textsubscript{pp} and $\Omega$ = $2\pi$ $\cdot$ 2.84~MHz with a DC bias of -20~V. A plot of the electric potential and pseudopotential in the radial plane is shown in Figure~\ref{fig:pseudopotential}. A plot of the pseudopotential in the radial and axial direction along a line through the origin and in between the trap rods is also shown, along with a quadratic fit. In the radial plane the computed pseudopotential fits well to a harmonic well, and is used to compute the transverse empty-trap secular frequency of 320~kHz. At the operating point, the Mathieu stability parameter is $q$ = 0.67, indicating stable ion trapping within the first stability region. A plot of the trap depth, Mathieu $q$ parameter, transverse secular frequency, and ratio of the drive frequency to secular frequency as a function of the drive voltage V\textsubscript{pp} and drive frequency $\Omega$ is shown in Figure~\ref{fig:contour}.

\section{Ion Production and Clock Operation}
\begin{figure}
\begin{center}
\leavevmode
\includegraphics[width=0.8\linewidth]{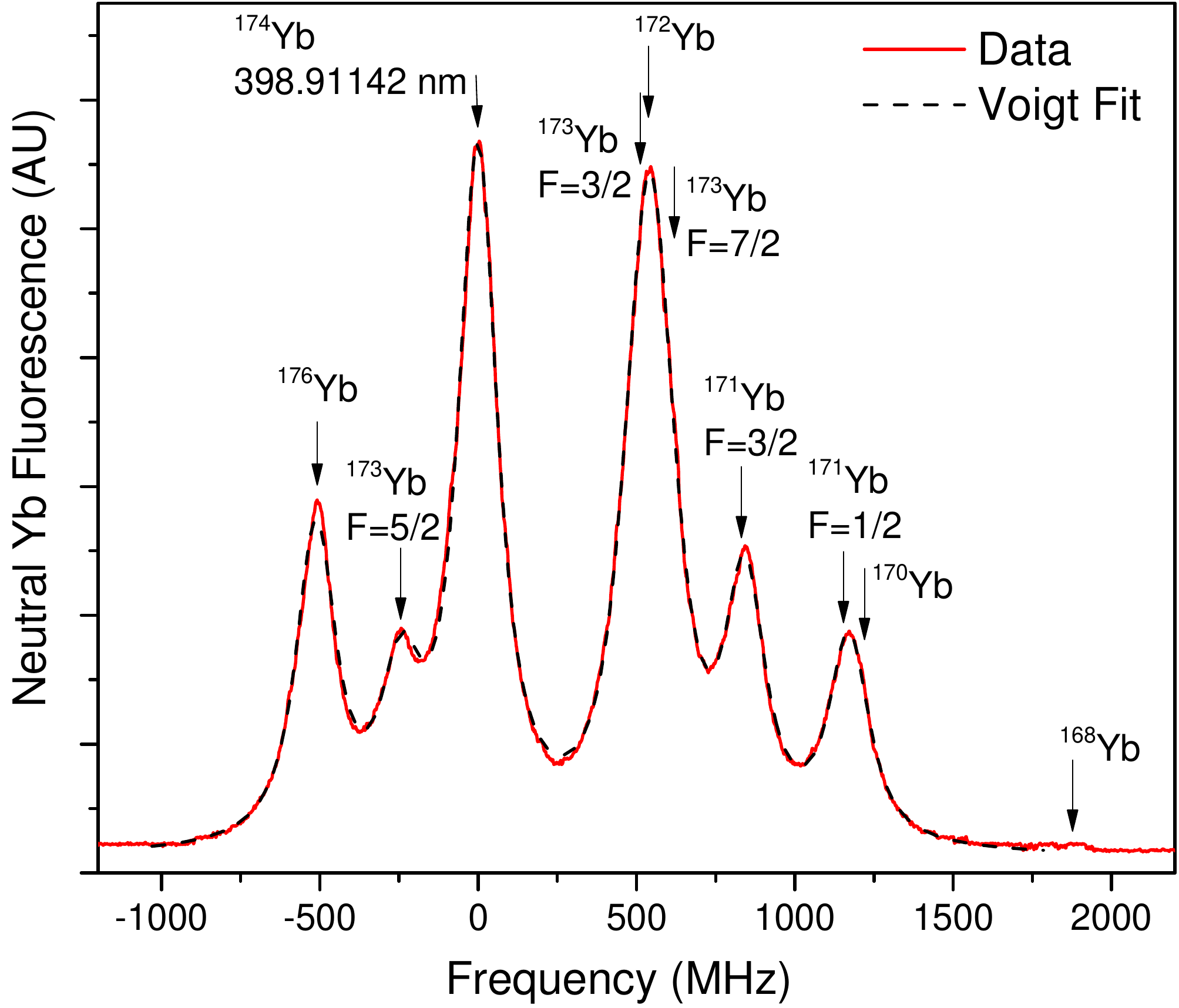}
\end{center}
\caption{Neutral spectrum of Yb from natural oven. The seven naturally occurring isotopes of Yb are visible, and each peak is fit to a Voigt profile representing the Doppler-broadened and power-broadened atomic transition. The composite Voigt fit is shown.} \label{fig:neutral}
\end{figure}

Ions are produced by resonant two-photon photoionization \cite{johanning, onoda} in the middle of the trapping region. The Yb oven is heated externally to a temperature of 500 \textdegree{}C, generating a broadly diverging neutral atomic beam in the trapping region. Operation of the natural oven while scanning the 399~nm laser around the vicinity of the neutral Yb resonances results in the neutral fluorescence spectrum shown in Figure~\ref{fig:neutral}, where all seven isotopes of Yb are visible. For ionization, a resonant 399~nm laser excites neutral Yb atoms from the \textsuperscript{1}S\textsubscript{0} state to the \textsuperscript{1}P\textsubscript{1} state. Subsequently, laser radiation with an energy greater than the ionization potential (wavelength less than 394~nm) results in ion production. Using this method, we can load either \textsuperscript{171}Yb\textsuperscript{+} or \textsuperscript{174}Yb\textsuperscript{+} in the trap with a loading time of several seconds, depending on the oven used and the frequency of the 399~nm laser radiation.

\begin{figure}
\begin{center}
\leavevmode
\includegraphics[width=0.8\linewidth]{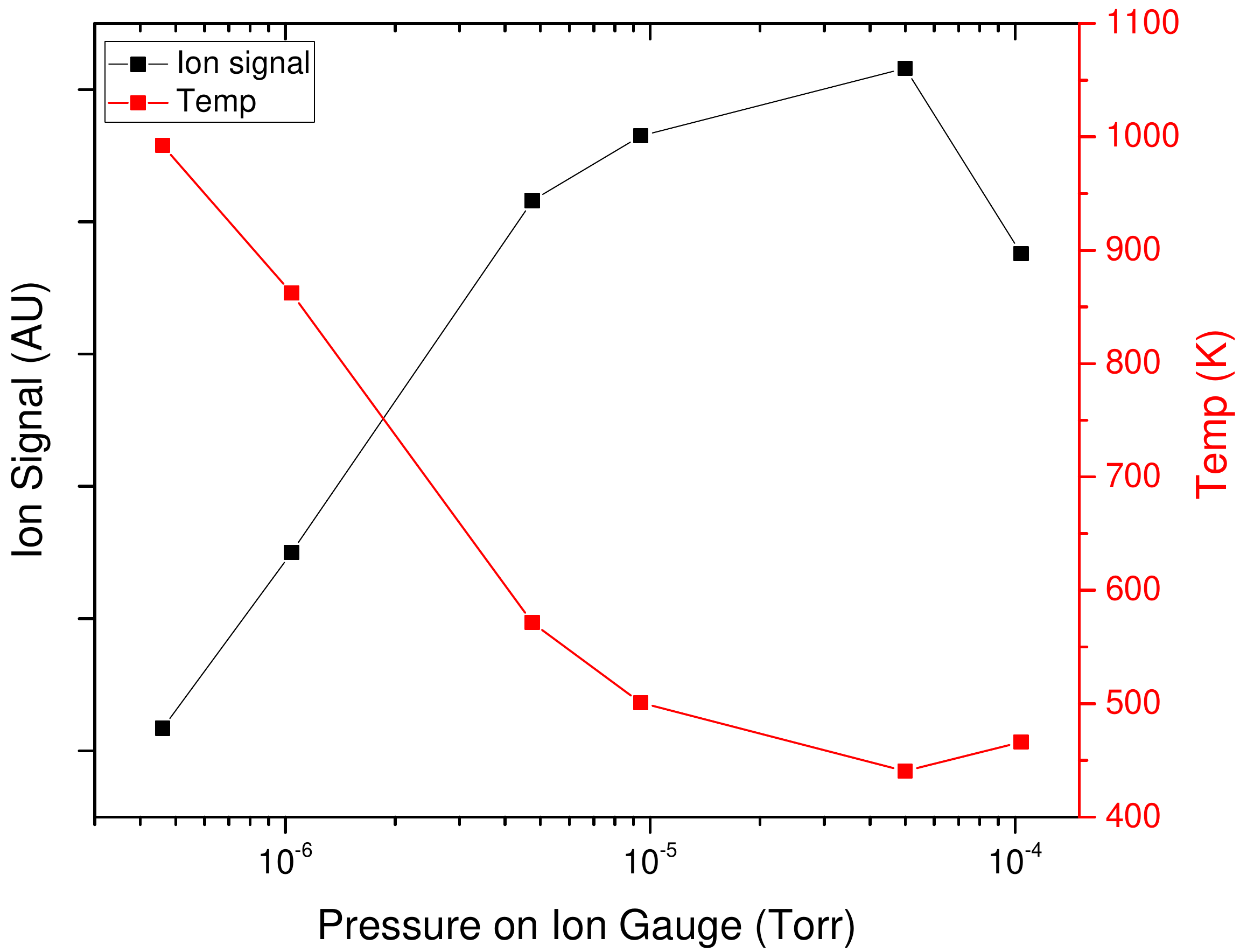}
\end{center}
\caption{Maximum ion count rate (AU) and Doppler Temperature vs ion gauge pressure.} \label{fig:doppler}
\end{figure}

Ions are cooled via collisional thermalization with Ne buffer gas, introduced into the trapping region via a manual leak valve in the vacuum chamber. In order the determine the optimum buffer gas pressure, the on-resonance ion fluorescence rate and Doppler width of \textsuperscript{174}Yb\textsuperscript{+} were recorded as a function of pressure on the ion gauge at a fixed Yb oven temperature,  shown in Figure~\ref{fig:doppler}. The on-resonance ion fluorescence peak occurs at the same point as the minimum Doppler temperature of 440~K, corresponding to a Doppler width of 930~MHz.

\begin{figure}
\begin{center}
\leavevmode
\includegraphics[width=1\linewidth]{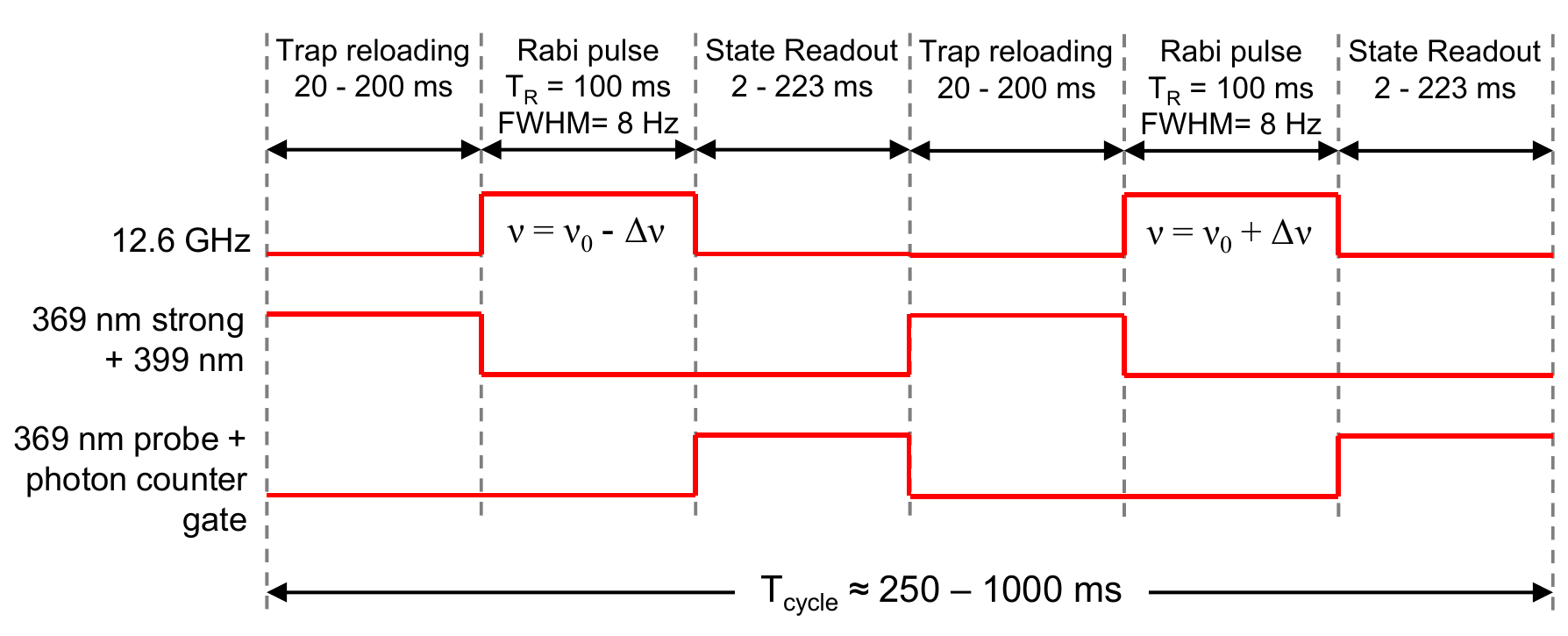}
\end{center}
\caption{Timing sequence for clock operation.} \label{fig:timing}
\end{figure}

Two collisionally-mediated pathways exist \cite{schauer} that allow for population transfer from the D-states to the long-lived F-state. While the F-state is not radiatively cleared in our experiment, it is collisionally depopulated, allowing F-state ions to slowly return to the cycling transition. Reduction in signal due to trap loss is mitigated by including a short (10-20~ms) trap reloading step in the cyclic timing sequence in order to maximize the number of ions contributing to the clock signal. This generic timing sequence is shown in Figure~\ref{fig:timing}. 

\begin{figure}
\begin{center}
\leavevmode
\includegraphics[width=0.85\linewidth]{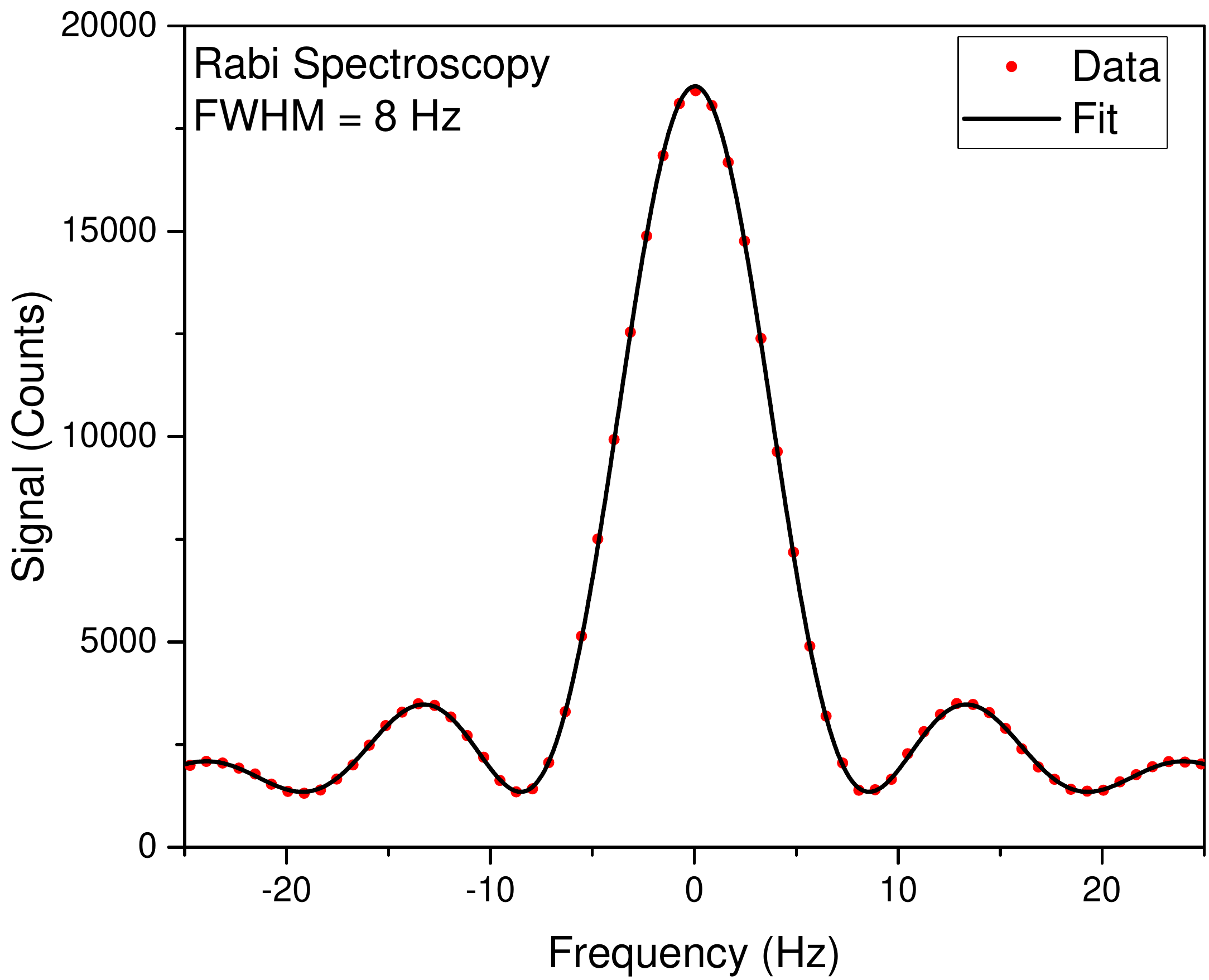}
\end{center}
\caption{Rabi spectroscopy with a 100 ms long $\pi$ pulse (5 averages, no background subtraction). The vertical axis shows the number of photon counts detected in the 13~ms photon counter gate time.} \label{fig:rabi}
\end{figure}

\begin{figure}
\begin{center}
\leavevmode
\includegraphics[width=0.85\linewidth]{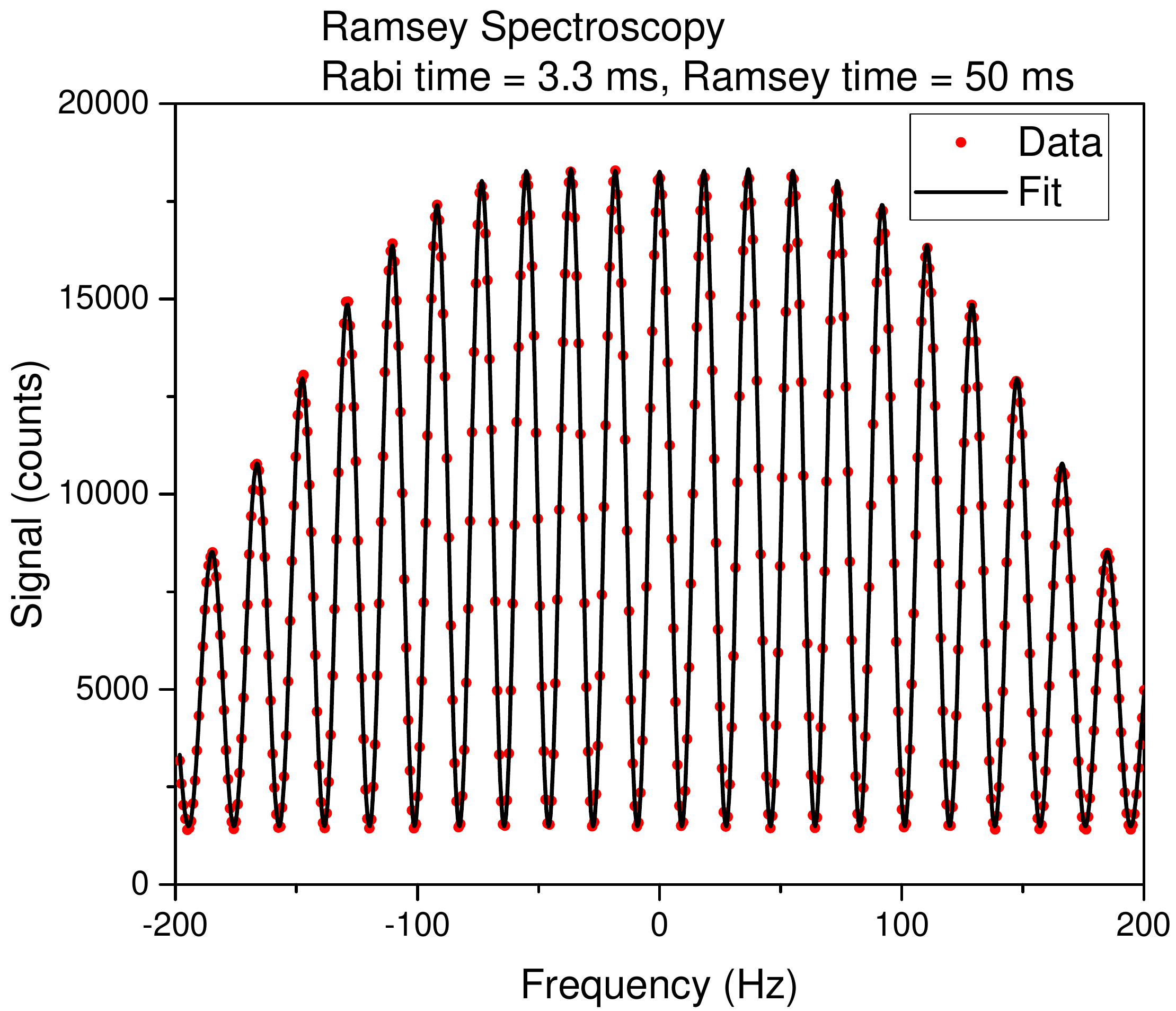}
\end{center}
\caption{Ramsey spectroscopy with a 50 ms long Ramsey free precession time (5 averages, no background subtraction). The vertical axis shows the number of photon counts detected in the 13~ms photon counter gate time.} \label{fig:ramsey}
\end{figure}

Optical pumping into the lower hyperfine ground state \mbox{occurs} via off-resonant excitation. The 369~nm optical pumping beam (which includes $\sigma$\textsuperscript{+}, $\sigma$\textsuperscript{-}, and $\pi$ polarization) excites ions into the \textsuperscript{2}P\textsubscript{1/2} $\mid$$F'$=1$>$ state, which allows for decay into the \textsuperscript{2}S\textsubscript{1/2} $\mid$$F$=0, $m\textsubscript{F}$=0$>$ state. Since the optical pumping beam is only 2.1~GHz detuned from this 930~MHz wide Doppler-broadened transition, it pumps ions into the lower clock state. Microwave spectroscopy is performed by applying a 440~mG C-field and scanning the 12.6~GHz microwaves across the \textsuperscript{2}S\textsubscript{1/2} $\mid$$F$=0, $m\textsubscript{F}$=0$>$ $\rightarrow$ $\mid$$F$=1, $m\textsubscript{F}$=0$>$ hyperfine ``clock" transition. A typical sequence for Rabi spectroscopy involves a 100-ms long microwave $\pi$ pulse, followed by state readout with the 369~nm probe beam. We observe an interaction-time limited FWHM of 8~Hz in the Rabi data shown in Figure~\ref{fig:rabi}. The results from Ramsey spectroscopy with a 50~ms long Ramsey free precession time are shown in Figure~\ref{fig:ramsey}.

\section{Clock Performance}
\begin{figure}
\begin{center}
\leavevmode
\includegraphics[width=0.9\linewidth]{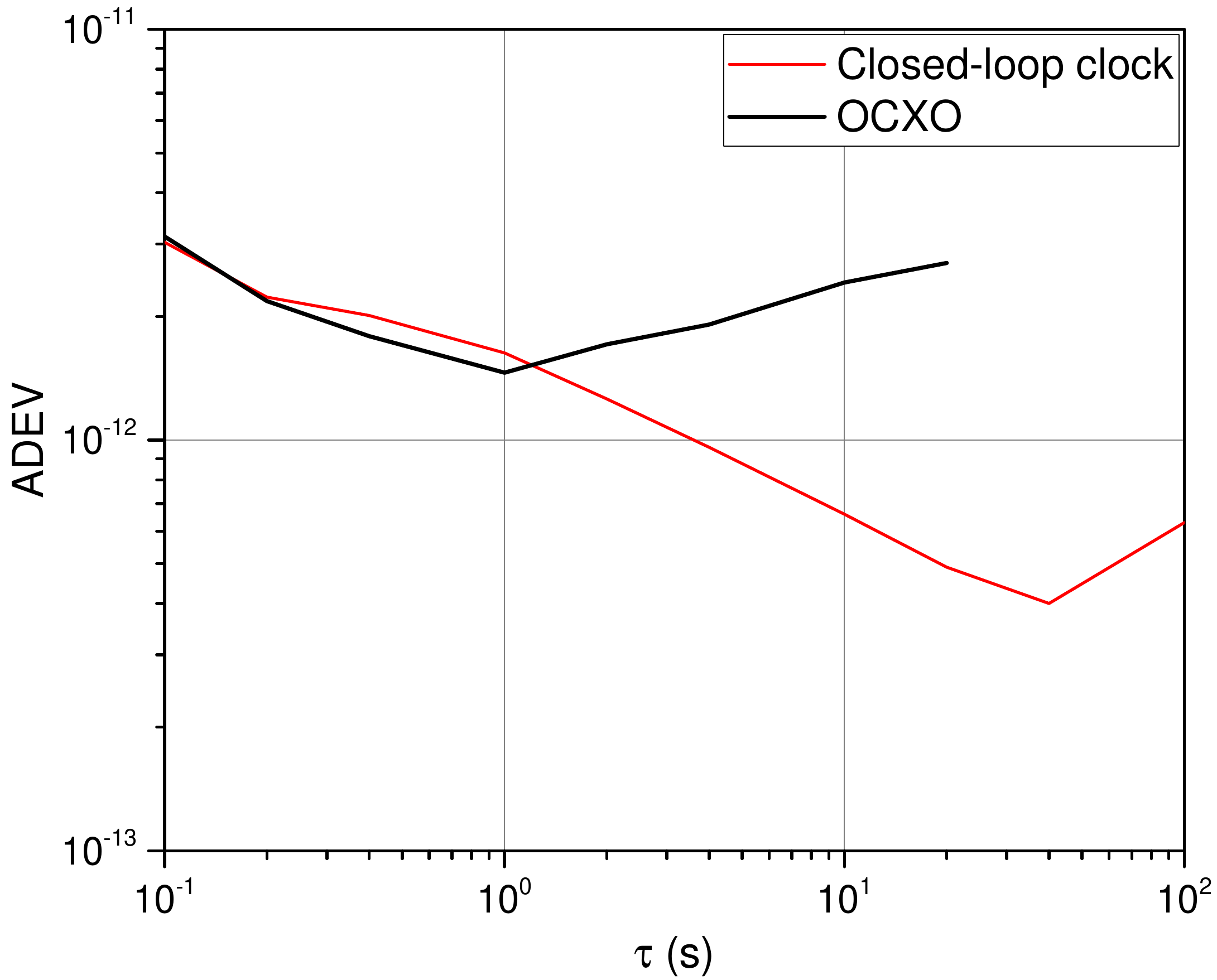}
\end{center}
\caption{Allan Deviation of free-running OCXO and closed-loop atomic clock based on Rabi interrogation time of 100~ms.} \label{fig:adev}
\end{figure}

Closed-loop clock operation is obtained via square-wave frequency modulation (FM) of the microwave interrogation frequency on either side of the atomic resonance, computation of an error signal, and subsequent steering of a voltage-controlled oscillator in a first-order frequency-locked loop \cite{cutler, vanier_fll}. The trap reloading time is 20~ms, the microwave interrogation time is 100~ms, and the photon counter gate time is 13~ms. The clock cycle time (total time to perform an interrogation on the left side of the line followed by an interrogation on the right side of the line) is 270~ms. The gain of the clock loop was set experimentally to respond to an impulse after $\sim$5 corrections, resulting in an ``attack time" on the order of 1~s. A plot of the measured closed-loop Allan Deviation $\sigma_y(\tau)$ as well as the open-loop OCXO performance is shown in Figure~\ref{fig:adev}. The closed-loop ADEV follows $1.6 \cdot 10^{-12} \tau^{-1/2}$ in the short term. The short-term stability of the OCXO ($1.5 \cdot 10^{-12}$ at 1~s) provides the largest contibution to clock short-term stability. The remainder is due to additional sources of noise, such as technical noise, photon shot noise, and quantum projection noise.

An estimate of the number of ions in the trap and the number of detected counts per ion at 369 nm can be performed as follows. For this estimate, we operate the natural oven and trap \textsuperscript{174}Yb\textsuperscript{+}. After first shelving ions in the D-state, we then clear the D-state using a resonant laser at 935~nm and monitor ion fluorescence at 297~nm, where each ion scatters one photon. The number of trapped ions can be determined based on the total ion fluorescence and the color filter transmission, geometrical solid angle, and PMT quantum efficiency. We estimate the solid angle to be 4\% to within a factor of 2, resulting in a total ion number of 20,000 with similar uncertainty. During normal operation with the enriched \textsuperscript{171}Yb oven while collecting flourescence at 369~nm, the total number of detected counts at the peak of the Rabi resonance is 17,200 counts, as shown in Figure \ref{fig:rabi}. The number of detected counts per ion is given by the ratio of the peak signal height to the total number of ions, yielding 0.9 counts per ion to within a factor of two. Because the number of detected counts per ion is close to unity, we operate in the limit where we have approximately equal contributions of photon shot noise and quantum projection noise \cite{santarelli}.



The Dick Effect describes the process by which microwave phase noise is aliased into the locked clock, limiting the frequency instability of a pulsed atomic clock \cite{dick1, dick2, santarelli_dick}. We compute the magnitude of the Dick Effect contribution based on a numerical calculation as follows. In the time domain, there exists a sensitivity function $g(t)$ which describes the time-dependent sensitivity of the final upper clock state population to instantaneous changes in microwave interrogation frequency during the Rabi pulse. For Rabi interrogation, the sensitivity function looks approximately like a half-sine function over the 100~ms Rabi interrogation time. In the frequency domain, this acts as a filter with a first null at approximately 15 Hz and a roll-off of 40 dB/decade. The phase noise of our microwave interrogation signal passes through this filter and is then converted to amplitude fluctuations via the static sensitivity of the slope of the Rabi resonance at the half-max (interrogation) point. The effects of frequency modulation on either side of the Rabi resonance are then taken into account. Subsequently, the digital sampling process results in aliasing of noise at higher frequencies into the first sampling zone. In our case, higher order terms result in negligible contributions to the overall effect. Synchronous demodulation and then decimation by a factor of two are then taken into account in our numerical model. Finally, the resulting amplitude fluctuations are passed through a numerical model of the single-pole clock loop filter, converted into a voltage correction on the OCXO with a resulting frequency noise, and finally converted to ADEV. The resulting Dick-effect limited ADEV for our microwave phase noise spectrum and Rabi interrogation parameters is $3.4 \cdot 10^{-13} \tau^{-1/2}$. In practice, the short-term performance of the OCXO ($1.5 \cdot 10^{-12}$ at 1 s) provides the largest limitation to the closed-loop ADEV in our apparatus.

\section{Conclusion}
The previous generation of miniature buffer-gas cooled trapped-ion clock development in \textsuperscript{171}Yb\textsuperscript{+} \cite{IMPACTAPL, IMPACT_fcs} was limited by significant laser scattering at 369~nm, requiring use of the 297~nm detection pathway, with photon shot noise limited \textit{SNR}. By operating the clock with a macroscopic light collection system, we have shown that it is possible to use direct detection at 369~nm. For the level of 369~nm laser scattering in our system, this improved the detection \textit{SNR} by a factor of 100. We have demonstrated that this 3~cm\textsuperscript{3} vacuum package is capable supporting a short-term stability of $1.6 \cdot 10^{-12}  \tau^{-1/2}$. 

\section{Acknowledgements}
We acknowledge assistance from John Prestage and Sang Chung of Jet Propulsion Laboratory in fabricating the ion trap vacuum package.



%

\end{document}